\documentclass[12pt]{article}

\usepackage[cp1251]{inputenc}
\usepackage[english,russian]{babel}
\usepackage{graphicx}
\usepackage{indentfirst}
\usepackage{epsfig,amsmath,amsfonts}

\inputencoding{cp1251}
\def\a{\alpha} \def\b{\beta} \def\g{\gamma} \def\d{\delta} \def\e{\epsilon}
\def\ve{\varepsilon}   \def\q{\theta}
   \def\l{\lambda} \def\m{\mu}
\def\n{\nu}  \def\p{\pi}  
 \def\s{\sigma}   \def\f{\varphi}
 \def\c{\chi} \def\y{\psi} \def\w{\omega}

   \def\D{\Delta}

  \def\P{\Pi}  
   
\def\F{\Phi}  \def\Y{\Psi} 
\def\DD{\lefteqn{D}\,/}
\def\dD{\lefteqn{\partial}/}

\def\fr{\frac} \def\dfr{\dfrac} \def\dt{\partial}

\def\beq{\begin{equation}}
\def\eeq{\end{equation}}

\def\bear{\begin{eqnarray}}
\def\eear{\end{eqnarray}}

\def\bea*{\begin{eqnarray*}}
\def\eea*{\end{eqnarray*}}

\def\ph{\phantom}

\def\leq{\lefteqn}

\def\bX{{\mathbf X}}
\def\bPi{{\mathbf \P}}
\def\bB{{\mathbf B}}
\def\bF{{\mathbf F}}

\begin{document}
\renewcommand{\figurename}{Fig.}
\renewcommand{\contentsname}{}
\renewcommand{\refname}{\begin{center}References\end{center}}

\title{Comments on QED with background
electric fields}\author{E.T. Akhmedov\footnote{akhmedov@itep.ru}{ } and
E.T. Musaev\footnote{musaev@itep.ru}\\
{B.Cheremushkinskaya, 25, ITEP, 117218, Moscow, Russia} \\ {and}
\\
{Moscow Institute of Physics and Technology, Dolgoprudny, Russia}}\date{}
\maketitle

\begin{center}Abstract\end{center}

It is well known that there is a total cancellation of the
\emph{factorizable} IR divergences in unitary interacting field
theories, such as QED and quantum gravity. In this note we show
that such a cancellation does not happen in QED with background
electric fields which can produce pairs. There is no factorization
of the IR divergences.

\section{Introduction}

The particle creation in external fields is among the most
interesting problems in quantum field theory. The effect of pair
creation in QED with external electric field was investigated from
different points of view in many places. The pair creation rate
was calculated in \cite{Schw}.

There are two reasons why we would like to address QED in electric
field back\-ground. The first one is that we would like to define
an appropriate setting to take into account the back-reaction of
the pair production on to the external field. The second reason of
considering the QED with background electric field is its
similarity with QFT on curved de Sitter background, which goes
beyond \cite{emil} the pair creation
\cite{Gibbons:1977mu},\cite{Mottola} and acceleration of
particles.

In particular, here we are interested in the IR behavior of QED
with electric field background. It is well known that there is
total cancellation of IR divergences in QED \emph{without}
background fields \cite{Weinberg}. The latter consideration can be
linked to the fact that mass-shell electrons can not radiate
mass--shell photons. In fact, consider the process
$e^-\rightarrow\g+e^{-*}$. Obviously the amplitude of this process
in the leading order is proportional to:

$$ A \propto \langle0,\mbox{out}|a_k^{-}\b_q^{-}
\int dt \, \hat{H}_{int}(t) \, a_p^{+}|0,\mbox{in}\rangle \propto
\int{d^4x} e^{i(p - q - k)x} \propto \delta^{(4)} \left(p - q -
k\right),$$ i.e. obviously there is the energy-momentum
conservation at the vertex in QED, if there are no any background
fields: $p=k+q$, where $p, q, k$ -- are momenta of the incoming
electron and outgoing photon and electron, respectively;
$\hat{H}_{int}$ -- is the QED interaction Hamiltonian describing
the interactions between electrons and photons. All of the three
legs of the amplitude are on-shell. Hence, $k^2-m^2= p^2 - m^2 =
0$ and $q^2 = 0$. Due to the latter relations the argument of the
$\delta$-function is never zero. Hence, the amplitude is zero,
which just means that there is no radiation on mass-shell.

However, if one of the particles is off-shell, say $k^2-m^2= \l$,
where $\l$ is the virtuality, then for the amplitude to be
non-vanishing it has to be that $\l=-2pq$. Such a dependence of
$\l$ on $q$ is important for the factorization of IR divergences,
which, in turn, is important for their cancellation to all orders
\cite{Weinberg}, \cite{Smilga:1985hp}. Note that such a relation
between the virtuality of the matter field and the momentum of the
radiated particle is a very special situation following from the
energy--momentum four--vector conservation at the vertices and
from the standard dispersion relations.

Let us sketch here the physical meaning of the cancellation of the
IR divergences. One can immediately notice that loop corrections
to any processes in QED have IR divergences, which are all of the
same order (independently of the number of loops) as the IR
cut-off parameter is taken to zero \cite{Weinberg},
\cite{Smilga:1985hp}. E.g. the first loop corrections have a
characteristic IR divergence as follows:
\beq\nonumber
\mbox{IR}_{\mbox{\scriptsize{loop}}}\propto\int\fr{d^4q}{(pq)^2q^2}
\propto\log{m_0}
\eeq
with the cut-off $m_0\rightarrow0$. Due to the factorization of the IR
divergences higher loops bring just powers
of such an expression  \cite{Weinberg}. Because of such contributions, if
the IR cut-off is taken to zero, all the cross--sections in QED
appear to be zero, which is quite puzzling.

The resolution of this problem comes with the understanding that
any scattering process of hard particles is accompanied with the
emission of the tree level \emph{soft} photons (because electrons
do accelerate during the scattering process) \cite{Weinberg}. As
the result the cross--sections of hard processes are dressed with
the powers (due to the factorization) of the contribution as
follows. The amplitude for the emission of a soft photon (with the
momentum $|\vec{q}|\to 0$) is proportional to the propagator of
the virtual particle, which, in its own right, is proportional to
its inverse virtuality $1/\l\propto 1/(pq)$. The latter behavior
of the propagator follows from the presence of the above mentioned
$\delta$--function (enforcing energy conservation) in the
interaction vertex.

Thus, after the integration over the invariant phase volume of the
emitted photon, the factor contributing to the cross-section is
proportional to \cite{Weinberg}:
\beq\nonumber
\int|M(k,q;p)|^2\fr{d^3q}{|\vec{q}|}
\propto\int\fr{1}{(pq)^2}\fr{d^3q}{|\vec{q}|}\propto\log{m_0}.
\eeq
Such contributions come exactly with the appropriate signs to
cancel the above men\-tioned loop IR divergences \cite{Weinberg}.
Higher loops are cancelled by multiple photon emissions. On the
other hand, from the very beginning one can dress electron legs
with soft photon legs and avoid IR divergences both in loops and
tree--level contributions \cite{BlochNordseik}.

The general goal of this note is to show that considering QFT in
background fields should lead to some problems through one or
another way, if the background field is taken into account as the
source and if one uses in calculations the exact matter harmonics
in the background fields (corresponding to the excitations above
false vacuum state) instead of the standard plain waves
(corresponding to the excitations above the correct vacuum). The
reason why we expect such problems is that, if the background
field is taken as classical and fixed (rather than as a quantum
state) and if one does not take into account back--reaction, this
makes the system non--closed and should lead to some
inconsistences.

In these notes we show that there is no cancellation of IR
divergences in QED with background electric field. It happens
because, due to the presence of back\-ground fields, we do not
have energy--momentum four--vector conservations at the vertices.
Of course the total energy of the background field and of the
particles participating into reactions is conserved, if one takes
into account the back--reaction.  However, if the background field
is considered as fixed, and one looks only at the four--momenta of
the particles participating into the reactions then these momenta
do not obey the energy conservation condition.

As the result, the virtuality of a matter field, participating
into a reaction, is not related to the momentum of the radiated
photon and there is no factorization of the IR divergences.
Moreover, the standard procedure a la Bloch and Nordsiek
\cite{BlochNordseik} does not apply to the case of the presence of
background fields. Concisely, the method works as follows. One
dresses external electron legs with soft photon field, which is
not capable of creating pairs. Corresponding vacuum polarization
operator is equal to one. However, in the presence of the external
field capable of creating pairs there is always vacuum
polarization, because in this case QED is build on the unstable
(false) vacuum constructed via creation and annihilation operators
of the exact harmonics in the background field. For the same
reason the Lee and Nauenberg theorem \cite{Lee:1964is} does not
apply in the case under consideration, because in the presence of
the background fields in--vacuum state does not coincide with the
out--vacuum.

Thus, we see that the standard methods of the cancellation of the
IR divergences do not work in background fields. If the IR
divergences do not cancel, then, as we discuss in the main body of
the paper, this should lead to many problems even at the leading
order. In particular, it could lead to such problem as the
presence of the infinite electron self--energy. We do not declare
that IR divergences definitely do not cancel, but obviously one
should take our observation as a problem, which should be resolved
somehow.

It can be shown \cite{Gitman1} that, if background electric field
creates a \emph{finite} number of pairs, then one can build a
unitary $S$--matrix in the theory. However, if the background
field is capable of creating infinite number of pairs, then the
corresponding evolution operator even is not unitary
\cite{Gitman1}. Such a situation for QED with background electric
fields is rather unphysical, because the background field should
contain infinite amount of energy to be able to create infinite
number of pairs. However, it is exactly this situation which one
encounters in a QFT in de Sitter space background \cite{emil},
because to maintain de Sitter isometry one has to consider the
space as being eternal and, hence, capable of creating infinite
number of pairs.

\section{Harmonics in pulse background}

In this section we examine the QED in the \emph{pulse} electric
field background:
\beq\label{1.1}
A_{\m}=(0;0,0,a\,\tanh \a
t),\quad{}\vec{E}=(0,0,\fr{a\a}{\cosh(\a{}t)^2}).
\eeq
Note that $|E|\to 0$, as $t\to \pm \infty$. Dirac equation is as
usual:
\beq\label{1.2}
(i \DD - m)\Y=0.
\eeq
Here the covariant derivative is: $D_{\m}=\dt_{\m}-i e A_{\m}$.

Solutions of this equation can be represented in the form:
\beq
\Y=(i\DD+m)\F,
\eeq
where $\F$ satisfies the equation, which is similar to the
Kl\"{e}in-Gordon one:
\beq
(\dt_{\m}\dt^{\m}-2ieA^{\m}\dt_{\m} -
e^2A_{\m}A^{\m}+m^2-ie\dt_{\m}A_{\n}\g^{\m}\g^{\n})\F=0.
\eeq
Since the operator $i\DD+m$ is twice degenerate we choose two
independent solutions:
\bea*
\F_1=\f_1R_1;\\
\F_2=\f_2R_2.
\eea*
Where $R_{1,2}$ are two eigenvectors of the matrix $\g^0\g^3$
which correspond to the eigenvalue $\l=+1$. In the standard
representation of gamma-matrices:
\beq
R_1=\left[
\begin{array}{ll}
0\\-1\\0\\1
\end{array}\right]\quad
R_2=\left[
\begin{array}{ll}
1\\0\\1\\0
\end{array}\right]
\eeq
These solutions will stay independent after the action of the operator
$i\DD+m$.

Thus, functions $\f_1$ and $\f_2$ satisfy the following equation:
\beq\label{1.3}
\left(\dt_{\m}\dt^{\m}+2i e \tanh\a t\,\dt_3+e^2a^2\tanh^2{\a
t}+m^2-\fr{iea\a}{\cosh^2{\a t}}\right)\f=0.
\eeq
We will look for the solutions of this equation in the following
form:
\beq
\f=\c_k(t)e^{-ik_ix^i},
\eeq
where $\c_k(t)$ satisfies:
\beq\label{1.4}
\ddot{\c_k}(t)+\left[\w^2(t)-\fr{i e a}{\cosh^2{\a
t}}\right]\c_k(t)=0.
\eeq
Here $\w^2(t)=k_1^2+k_2^2+(k_3+ea\tanh(\a{}t))+m^2$.

Positive energy solutions at the past infinity ($t \to - \infty$)
have the following form \cite{Grib}:
\beq \label{1.5} \c_k^{+}(t)=e^{-i 2\a\m \,t}(1+e^{2\a t})^{-i\q}
F[\b,\g;\d;-e^{2\a\,t}]=(\c_{-k}^{-}(t))^*, \eeq where
\bea*
&&\b=-i\q-i\m-i\n;\quad \g=-i\q-i\m+i\n;\\
&&\d=1-2i\m;\quad
\q=\fr{ea}{\a};\\
&&2\a\m=\sqrt{k_1^2+k_2^2+(k_3-ea)^2+m^2};\\
&&2\a\n=\sqrt{k_1^2+k_2^2+(k_3+ea)^2+m^2}.
\eea*
Solutions of the Dirac equation are:
\beq \label{1.6}
\y^{\pm}_{r,k} = (i\g^0\dt_0\pm\g^ik_i+e\g^{\m}A_{\m}+
m)\c^{\pm}_k(t)e^{\mp{}ik_ix^i}R_r=:f^{\pm}_{r,k}e^{\mp{}ik_ix^i}.
\eeq
Asymptotics of the functions $\c_{k}^{\pm}(t)$ are \cite{BE}:
\bea*
&&\c^{+}_{k}(t)\stackrel{t\rightarrow-\infty}\longrightarrow{e^{-i\w_{-}t}};\\
&&\c^{-}_{k}(t)\stackrel{t\rightarrow-\infty}\longrightarrow{e^{+i\w_{+}t}},
\eea*
where $\w_{\pm}=\lim_{t\rightarrow\pm\infty}\w(t)$. We see, that
spinors (\ref{1.6}) have the right asymptotics in the past to be
the definite energy solutions: $\y_{r,k}^{\pm}=(\pm
\lefteqn{k}/+m)R_{r}e^{\mp{}ikx}$, where $kx=k_{\m}x^{\m}.$

The usual scalar product of the two solutions in question is:
\bear\label{1.7}
\nonumber
\int{{\rm
d}^3x\,{\y^{\pm}_{r,k_1}}^{\dagger}\y^{\pm}_{s,k_2}}&=&\left[{\dot{\c}_k}^{*}\dot{\c}_k-i(\pm{}k_3+eA_3)\c^{*}_{k}\overleftrightarrow{\dt_0}{\c}_k+\w^2(t)\c^{*}_{k}\c_k\right]\times\\
&&\times2\,\d(\vec{k}_1-\vec{k}_2)\d_{rs}=4\w_{\mp}(\w_{\mp}+k_3\mp{}ea)\,\d(\vec{k}_1-\vec{k}_2)\d_{rs},
\eear
where $\c_{k}$ denotes, for short, $\c^{+}_k(t)$ or $\c^{-}_k(t)$
for $\pm$-energy solutions respectively.

Finally, with the normalization (\ref{1.7}) the general solution of the Dirac equation
in the external electric field in question can be written as:
\beq\label{1.8}
\Y(x)=\sum_{r}\int{\rm{}d}^3k\left[\fr{1}{\sqrt{2\w_{-}}}\Y^{+}_{r,k}a^{-}_{r,k}+\fr{1}{\sqrt{2\w_{+}}}\Y^{-}_{r,k}b^{+}_{r,k}\right],
\eeq
where
\beq \label{1.9}
\Y^{\pm}_{r,k}=\fr{1}{\sqrt{2}}(\w_{\mp}+k_3\mp{}ea)^{-1/2}\y^{\pm}_{r,k},
%
\eeq
$a^{-}_{r,k}$ ($b^{-}_{r,k}$) are annihilation operators of
particles (antiparticles) with spin index $r$ and momentum
$\vec{k}$.

Now we can define the ``in'' vacuum state $|0,\mbox{in}\rangle$
as: $a^{-}|0,\mbox{in}\rangle=b^{-}|0,\mbox{in}\rangle=0$. The
name for the state follows from the fact that the solution
(\ref{1.8}) consists of the in-harmonics, which behave as
solutions of free Dirac equation with \emph{definite energies}
only as $t\rightarrow-\infty$. Hamiltonian has the following form
\cite{Grib}:
\beq
\mathcal{H}=\int {\rm d}^3k\,\w_k(t)\,
\left[E_k(t)(a_k^{+}a_k^{-}-b_k^{-}b_k^{+}) + F_k(t)a_k^{+}b_k^{+}
+ F^{*}_k(t)b_k^{-}a_k^{-}\right],
\eeq
where $E_k(t)$ and $F_k(t)$ are constructed from the in-harmonics.
It can be seen that as $t\rightarrow\-\infty$,
$E_k(t)\rightarrow\mbox{const}$, $F_k(t)\rightarrow0$. The
in-vacuum $|0,\mbox{in}\rangle$ is not an eigenvector of this
Hamiltonian at general values of $t$, which is directly related to
the vacuum instability and pair creation. Note that the
Hamiltonian under consideration is time dependent, because there
is the time dependent background field. Hence, the energy is not
conserved and the system in question is not a Hamiltonian one.
However, we call the operator in question as the Hamiltonian
because, using its $T$-ordered exponent in the second quantized
formalism, we can build the Green function, which allows to
construct the solutions of the corresponding Dirac equation
(\ref{1.2}). I.e. the latter Green function describes the time
evolution in the system of the free fields.

To diagonalize this Hamiltonian at $t\rightarrow+\infty$ (where
$F_k(t)\neq0$) one should consider Bogolyubov transformations
\cite{Grib}:
\bea*
&&a_k^{-}=\a_k\tilde{a}_k^{-}+\b_k\tilde{b}_k^{+};\\
&&b_k^{-}=\a_k\tilde{b}_k^{-}-\b_k\tilde{a}_k^{+};
\eea* here the operators with the tilde are the creation and
annihilation operators for out-harmonics: out-harmonics are
defined to be free definite energy spinors at future infinity,
i.e. as $t\rightarrow+\infty$. As the result we have such a
situation that $|0,\mbox{in}\rangle \neq ({\rm phase})
|0,\mbox{out}\rangle$ \cite{Grib}, unlike the case of QED without
background fields.

Now we would like to address the question of whether the on-shell
electron (cor\-res\-pon\-ding to the exact solution of the Dirac
equation in the background field) can radiate photon or not. On
general physical grounds one can definitely give the answer
``yes'' on this question, because electrons will accelerate under
the action of the background field.

But let us see formally how the things work. The problem is that
due to the pair production in the background field it is hard to
define what do we mean by the $S$--matrix and the amplitude. The
photon is defined uniquely because it doesn't interact with
external field, but there are problems with electrons. In the
papers \cite{Nikishov,Gitman2} the $S$--matrix was constructed for
the case of the background fields.

Let us consider the amplitude of the process where
electron with momen\-tum $p$ radiates photon with momentum $q$:
\beq\label{1.9.1}
\langle0,\mbox{out}|\tilde{a}_k^{-}
\beta_q^{-}\left(\int{}d^4x\bar{\Y}\lefteqn{A}{\,/}\Y\right)a_p^{+}|0,\mbox{in}\rangle,
\eeq
where $\beta_k^{-}$ -- is the photon annihilation operator,
$\langle0,\mbox{out}|$ -- is the out vacuum state, which is
defined as
$\langle0,\mbox{out}|\tilde{a}^{+}_k=\langle0,\mbox{out}|\tilde{b}^{+}_k=0$.

We now write $\bar{\Y}$ and $\Y$ in eq.(\ref{1.9.1}) in terms of
``out'' and ``in'' harmonics, res\-pec\-tively. After some simple
transformations one obtains \cite{Grib,Nikishov,Gitman2}:

\bear\label{1.9.2}\nonumber
&\langle0,\mbox{out}|\tilde{a}_k^{-}\b_q^{-}\left(\int{}d^4x\bar{\Y}\lefteqn{A}{\,/}\Y\right)a_p^{+}|0,\mbox{in}\rangle=\langle0,\mbox{out}|0,\mbox{in}\rangle\int{d^4x}\tilde{\bar{\Y}}_k^{+}\ve_{\m}^{*}\g^{\m}e^{iqx}\Y_p^{+}+\\\nonumber
&+\int{d^4x}\int\dfr{d^3k_1}{\sqrt{2k_1^0}}\b^*_{k_1}\langle0,\mbox{out}|\tilde{a}_{k_1}^{-}a_p^+|0,\mbox{in}\rangle\tilde{\bar{\Y}}_k^{+}\ve_{\m}^{*}\g^{\m}e^{iqx}\Y_{k_1}^{-}+\\
&+\int{d^4x}\int\dfr{d^3k_1}{\sqrt{2k_1^0}}\b_{k_1}\langle0,\mbox{out}|\tilde{a}_{k}^{-}a_{k_1}^+|0,\mbox{in}\rangle\tilde{\bar{\Y}}_{k_1}^{-}\ve_{\m}^{*}\g^{\m}e^{iqx}\Y_{p}^{+}+\\\nonumber
&+\int{d^4x}\int\dfr{d^3k_1d^3k_2}{2\sqrt{k_1^0k_2^0}}\langle0,\mbox{out}|\tilde{a}_{k}^-\tilde{b}_{k_1}^-b^+_{k_2}a^{+}_p|0,\mbox{in}\rangle\tilde{\bar{\Y}}^-_{k_1}\ve^*_{\m}\g^{\m}e^{iqx}\Y^-_{k_2}.
\eear
The first term in the sum on the RHS of (\ref{1.9.2}) corresponds,
up to the factor $\langle0,\mbox{out}|0,\mbox{in}\rangle\neq 1$,
to the usual amplitude of the photon radiation. The other terms
appear because ``out'' and ``in'' vacuum states are not the same.
These terms (and the factor
$\langle0,\mbox{out}|0,\mbox{in}\rangle$ in the first term)
describe the pair creation by external field.

The tree level amplitude (\ref{1.9.2}) is divergent due to
infinite range of time integration. One can compute the
corresponding cross--section, after regularization, using the
optical theorem \cite{Nikishov,Gitman2}. However, to understand
the issue of the IR divergences, one needs to deal somehow with
the amplitudes themselves rather than with the cross--sections.
What can be done in such circumstances? Let us consider one of the
terms in (\ref{1.9.2}) which corresponds to the classical
radiation process. It has a characteristic contribution which
appears in all four terms in (\ref{1.9.2}) and leads to
\emph{non--factorizable} IR divergences.

 If one would consider the classical limit of the
amplitude (\ref{1.9.2}), then only some part of the first term
will survive: the one which is not sensitive to the change of the
vacuum. In fact, to define the classical amplitude one should
consider correlation function with three retarded Green
functions\footnote{Two retarded Green functions in the external
field for the incoming and outgoing electron legs, and the third
one --- for the outgoing photon.}, then amputate the external legs
and substitute them by the mass-shell exact harmonics. The
retarded Green functions are classical objects: these functions
are not sensitive to the choice of the vacuum because they are
derived from the $c$--numbered commutators of the fields. It is
worth stressing here that after the amputation of the external
retarded propagators we still have an ambiguity in the choice of
which type of the free harmonics we should substitute instead of
the propagators: everywhere in-- or out--harmonics, or
in--harmonics for the incoming waves, while out--harmonics for the
outgoing ones. The point is that the conceptual conclusions about
the possibility of the radiation on mass--shell do not depend on
what kind of harmonics we will choose.

Thus, the classical amplitude in question, which is responsible
for the description of the radiation process on mass--shell, is
proportional to:
\beq
M(k,q;p)\propto\int{}d^4x\overline{\Y^{+}_{s,k}}\g^{\m}\Y^{+}_{r,p}\e_{\m}^{*}e^{iqx},
\eeq
where $\Y_{s,k}^{\-}$ is given in (\ref{1.9}) and (\ref{1.6}).

Using (\ref{1.9}) we can write:
\bear\label{1.10}\nonumber
&&M(k,q;p)\propto\int{}\fr{d^4x}{(p_0+p_3-e{}a)^{1/2}(k_0+k_3-e{}a)^{1/2}}\bar{\y}^{+}_{s,k}\g^{\m}\y_{r,p}^{+}\e^{*}_{\m}e^{iqx}=\\
&&\int{}\fr{dt}{(p_0+p_3-e{}a)^{1/2}(k_0+k_3-e{}a)^{1/2}}\bar{f}^{+}_{s,k}\g^{\m}f_{r,p}^{+}e^{iq_0t}\e^{*}_{\m}\d(\vec{p}-\vec{k}-\vec{q}),
\eear
where $p_0=\w(p)_{-}$ and $k_0=\w(k)_{-}$.

Photon polarization vectors in Coulomb gauge are:
$\e^{\m}=\d^{\m}_1\pm\d^{\m}_2$. Taking this fact into account we
have:
\bear\label{1.11}\nonumber
&\bar{f}^{+}_{r,k}\g^{\m}f^{+}_{s,p}=&2\left\{\dot{\c}^{*}_k\dot{\c}_p+i\dot{\c}^{*}_k{\c}_pp_3+i{\c}^{*}_k\dot{\c}_pk_3\right.\\
&&\left.+{\c}^{*}_k{\c}_p[m(k_3-p_3)+(k_{+}p_{+}-k_3p_3)]\right\}(\d^{\m}_1+\ve\d^{\m}_2),
\eear
where $\ve=r-s=\pm1$. We see that photon can be radiated only if
the spin of the electron has been changed. Obviously this
fact is related to the spin projection conservation.

The integral (\ref{1.10}) is divergent. After regularization, we
can apply numerical methods to compute the integral in
(\ref{1.10}). We have com\-puted it with several different finite
limits of integration. The result of integration weakly depends on
the change of the integration limits.

\begin{figure}
\includegraphics[height=8cm,keepaspectratio]{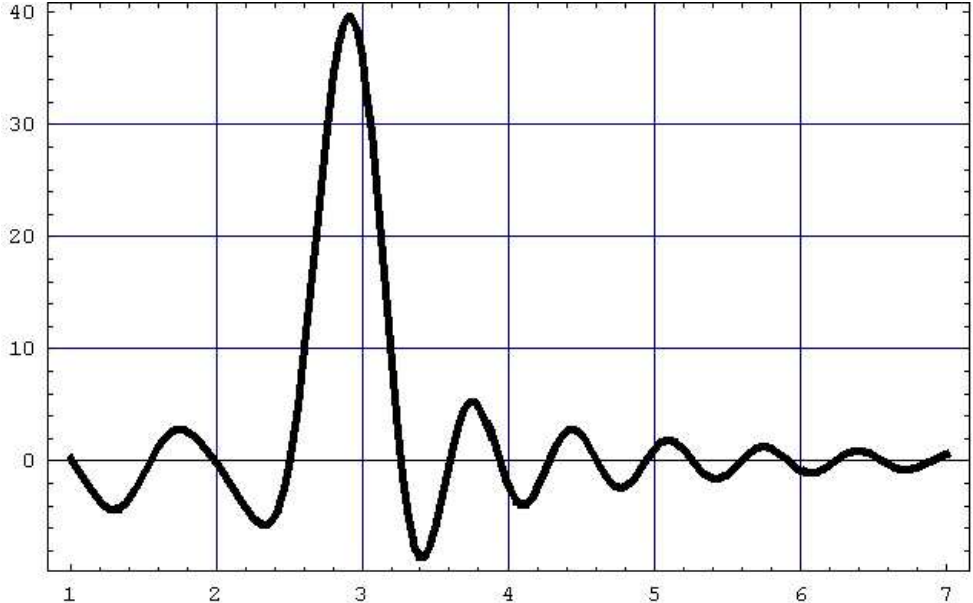}
\caption{Real part of the amplitude $e\rightarrow\g+e$ for the
pulse background} \label{fig1}
\end{figure}

Thus, numerical calculations show that unlike the case of QED
without background fields, the amplitude (\ref{1.10}) is not zero
if $\vec{p}=\vec{k}+\vec{q}$. This just ensures the obvious fact
that electron accelerates under the action of the background field
and emits photons.

The fact that the amplitude is not zero on mass--shell gives us a
gauge invariant way of answering affirmatively on the question
about the radiation on mass--shell, which was formulated before
the equation (\ref{1.9.1}). The Fig.\ref{fig1} shows the
dependence of the real part of the integral (\ref{1.10}) on $k_0$
for some fixed values of the other variables. It doesn't matter
what are their concrete values. For the concreteness we take
$p_0=4$ and $q_0=1$. Compare this discussion with the one in the
introduction.

It is worth stressing here that the amplitude (\ref{1.10}) is not
only non--zero on mass--shell, but as well is complex rather than
pure imaginary, as it should be at the leading order in any theory
with the unitary evolution operator. Hence, this fact rises the
question to \cite{Gitman1} on how in detail the unitarity
condition is ensured after the regularization of the IR
divergences.

Now if in the amplitude like (\ref{1.10}) one of the electron
legs, say that one, which is with the momentum $k$, will be
off-shell, then its virtuality $\l$ will not depend on the
momentum $q$ of the soft photon, because there is no simple
$\delta$--function ensuring energy conservation in the
corresponding vertex. The same is true for the total amplitude
(\ref{1.9.2}). Note that according to the Fig. \ref{fig1} the
delta function which was present in the vertices in the theory
without background fields is now smeared and as the result the
virtuality $\l$ of the matter particle is not directly related to
the momentum $q$ of the photon. This shows that there is no
factorization of the IR divergences in the conditions under study.

\section{IR divergences in loops}

In this section we show what kind of consequences one obtains even
at the leading order in the case if IR divergences do not cancel
out. To see them, let us derive the electron Green function in the
external field. Green function for the field $\Y(x)$ satisfies to
the following equation:
\beq\nonumber
(i\dD+e\lefteqn{A}\,/+m)G(x_2,x_1)=\d(x_2,x_1).
\eeq
Or in the formal operator representation:
\beq\label{2.2}
G=(\lefteqn{\P}/+m)^{-1},\quad{}\leq{\P}/=\leq{p}\,/+e\lefteqn{A}\,/.
\eeq
Here we understand Green function $G$ as an operator acting on the
states $|x\rangle$, $p$ is the usual momentum operator
\cite{Schw}. So the Green function is
$G(x_2,x_1)=\langle{}x_2|G|x_1\rangle$.

We can write $G$ in the integral form:
\beq\label{2.3}
G=-i\int_0^{\infty}ds(\leq{\P}/-m)\exp\left[i(\leq{\P}/\,^2-m^2)s\right].
\eeq
Introducing the unitary evolution operator
$U(s)=\exp(-iHs)=\exp(i\leq{\P}/\,^2s)$, we can write the Green
function as:
\beq\nonumber
G(x_2,x_1)=-i\int_0^{\infty}dse^{-im^2s}\langle{}x_2(0)|\leq{\P}/\,(s)-m|x_1(s)\rangle,
\eeq
where $|x(s)\rangle=U(s)|x(0)\rangle$.

Evolution of the operators $\P_{\m}$ and $x_{\m}$ is as follows:
\beq\label{2.4}\left\{
\begin{array}{lll}
\dfr{dx_{\m}}{ds}=&-i[H,x_{\m}]&=2\P_{\m};\\\\
\dfr{d\P_{\m}}{ds}=&-i[\P_{\m},H]&=-2eF_{\m\n}\P^{\n}+ie\dt^{\a}F_{\a\m}-\dfr{1}{2}\dt_{\m}F^{\a\b}\s_{\a\b},
\end{array}
\right.\eeq
where $F_{\m\n}=\dt_{[\m}A_{\n]}$ -- is the field strength tensor and
$\s_{\a\b}=i[\g_{\a},\g_{\b}]/2$.

For the simplicity we will use the matrix notation
($\bX=\|x_{\m}\|,\,\bPi=\|\P_{\m}\|\mbox{ etc.}$) for the
operators:
\beq\label{2.5}\left\{
\begin{array}{lll}
\dot{\bX}&=2\bPi;\\
\dot{\bPi}&=-2e\bF\bPi-\bB.
\end{array}
\right.\eeq
Now we can write the solution of these equations in symbolic form.
We will separate sectors $(0,3)$ and $(1,2)$ because matrix $\bF$
has only $F^0_{3}$ and $F^3_0$ nonzero components. From now on we
will understand the background value of $\bF$ as $2\times{2}$
matrix with components only in $(0,3)$-sector:
\beq\bF=\left[
\begin{array}{lll}
0&F^0_3\\
F^3_0&0
\end{array}
\right].\eeq Using these notations, we can write our symbolic
solutions as:
\bear\label{2.6}\nonumber
&&(0,3)\mbox{-sector}\\\nonumber
&&\bPi(s)=e^{-2e\bF{}s}\bPi(0)+\fr{1}{2e}\bF^{-1}\bB;\\\nonumber
&&\bX(s)-\bX(0)=\fr{1}{e}\bF^{-1}(1-e^{-2e\bF{}s})\bPi(0)+\fr{1}{e}\bF^{-1}\bB{}s;\\&&\\\nonumber
&&(1,2)\mbox{-sector}\\\nonumber &&\bPi(s)=\bPi(0);\\\nonumber
&&\bX(s)-\bX(0)=2\bPi(0)s.
\eear
Furthermore, the transformation function
$\langle{}x_2(0)|x_1(s)\rangle$ can be characterized by the
following equations:
\bear\label{2.7}\nonumber
i\dt_s\langle{}x'(0)|x(s)\rangle&=&\langle{}x'(0)|H|x(s)\rangle;\\
(i\dt_{\m}+eA_{\m}(x))\langle{}x'(0)|x(s)\rangle&=&\langle{}x'(0)|\P_{\m}(s)|x(s)\rangle;\\\nonumber
(-i\dt'_{\m}+eA_{\m}(x'))\langle{}x'(0)|x(s)\rangle&=&\langle{}x'(0)|\P_{\m}(0)|x(s)\rangle.
\eear
The boundary condition is:
$\lim_{s\rightarrow0}\langle{}x'(0)|x(s)\rangle =\d(x'-x)$.

To solve the first equation in (\ref{2.7}) we need $\bPi^2$ and $[\bX(s),\bX(0)]$:
\bear\label{2.8}\nonumber
&&(0,3)\mbox{ sector}:\\\nonumber
&&\bPi^2=(\bX(s)-\bX(0))\mathbb{K}(\bX(s)-\bX(0))+\fr{\bB{}s^2\bB}{\sinh(e\bF{}s)^2}\\\nonumber
&&+\fr{e\bB\bF}{\sinh(e\bF{}s)^2}(\bX(s)-\bX(0))+\fr{\bB{}\tanh(e\bF{s})}{2}\left(1+\fr{s}{e\bF}\right)\bB,\\\nonumber\\
&&[\bX(s),\bX(0)]=i\fr{1-e^{-2e\bF{s}}}{e\bF};\\\nonumber
&&(1,2)\mbox{ sector}:\\\nonumber
&&\bPi^2=(\bX(s)-\bX(0))\fr{1}{4s^2}(\bX(s)-\bX(0)),\\\nonumber\\\nonumber
&&[\bX(s),\bX(0)]=2is.
\eear
Here $\mathbb{K}:=e^2\bF^2\sinh(e\bF{}s)^{-2}/4$.

Now we can combine both sectors:
\bear\label{2.9}\nonumber
&\dfr{\langle{}x_2(0)|H|x_1(s)\rangle}{\langle{}x_2(0)|x_1(s)\rangle}=&(\bX_1-\bX_2)\tilde{\mathbb{K}}(\bX_1-\bX_2)+\fr{es}{\sinh(eEs)^2}\bB(\bX_1-\bX_2)+\\
&&+\fr{1}{2}\bB\tanh(eEs)\bF(1+\fr{s}{e\bF})\bB+\\\nonumber
&&+\fr{i}{2s}\mbox{Tr}\D_2+ie\coth(eEs)\mbox{Tr}\D_1+\fr{2s^2}{\sinh(eEs)^2}\bB\bF^{-1}\bB,
\eear
where $\tilde{\mathbb{K}}:=\mathbb{K}+\D_2/4s^2$ and
\beq
\D_1=\left[\begin{array}{llll}
1&0&0&0\\
0&0&0&0\\
0&0&0&0\\
0&0&0&1\\
\end{array}\right],\quad
\D_2=\left[\begin{array}{llll}
0&0&0&0\\
0&1&0&0\\
0&0&1&0\\
0&0&0&0\\
\end{array}\right].
\eeq
For our purposes we need only terms which dominate in the IR
limit $x^0_1-x_2^0\rightarrow\infty$.

The solution of the equation (\ref{2.7}) is:
\bear\label{2.10}
&\langle{}x_2(0)|x_1(s)\rangle\simeq&\exp\left[-\fr{i}{4}(\bX_1-\bX_2)(eE\coth(eEs)\D_1-\fr{1}{s}\D_2)(\bX_1-\bX_2)\right.\\\nonumber
&&\left.-\fr{i}{E}\left(s\coth(eEs)-\fr{1}{eE}\log[\sinh(eEs)]\right)\bB(\bX_1-\bX_2)\right].
\eear
As $\D{}x_0=x_1^0-x_2^0\rightarrow\infty$ the Green function behaves as:
\bear\label{2.11}\nonumber
&G(x_1-x_2)\propto&-i\int_0^{\infty}dse^{-im^2s}\left[\left(\fr{E\coth(eEs)}{2}\g^0+\fr{1}{2}\g^3\right)\D{}x_0\right]\times\\
&&\exp\left[-\fr{i}{4}eE\coth(eEs)\D{}x_0^2-\fr{i}{E}\left(s\coth(eEs)\ph{\fr{\log(\sinh(eEs))}{eE}}\right.\right.\\\nonumber
&&\left.\left.-\fr{\log(\sinh(eEs))}{eE}\right)e\dt_0F_{03}\g^0\g^3\D{x_0}\right].
\eear
Numerical calculations show that the Green function (\ref{2.11})
for the fermion in the back\-ground of the pulse electric field
(\ref{1.1}) is divergent as $\D{x_0}\rightarrow\infty$.

There are several points, which are worth stressing at this point.
First, the Feynman propagator, and both the Green
functions for in--in and in--out formalisms, in the external
electric field have as well similar to (\ref{2.11}) characteristic
divergence as $\D{x_0}\rightarrow\infty$. Second, as is well
known, the Green function for the fermions in the theory without
external fields is vanishing as $\D{x_0}\rightarrow\infty$. Third,
if the external field is magnetic, Green function as well is vanishing in the limit
$\D{x_0}\rightarrow\infty$.

Because of the divergence of the Green function in question one
can straightforwardly show that even the first loop diagrams, in
the QED with the background field under consideration, do have IR
divergences. E.g. even the electron self-energy diagram does have
IR divergence. It is worth stressing that without background
fields electron self--energy is IR finite.

 The divergence in question means that renormalized electron mass
in the background electric field is infinite if the IR cutoff is
taken to zero, which shows that such a formulation of the second
quantized interacting field theory in the electric field
background has some unavoidable problems.

One could hope to cancel these loop IR divergences by dressing
electron legs with photon legs \cite{BlochNordseik} or by using
observations of the Lee and Nauenberg \cite{Lee:1964is}. However
these procedures work only in the case when (see e.g. book of
Bogolyubov and Shirkov)

\bear \left\langle out \left| \hat{S} \right| in \right\rangle =
1, \eear but in our case this does not happen because $|in \rangle
\neq |out \rangle.$ So we clearly see problems with IR divergences
in the standard formulation of QFT in background fields. Similar
conclusions one can make for the QED in the constant (in space and
time) background electric field. Apparently the situation is
similar to the one with QFT on de Sitter space background
\cite{emil}, which is, in particular, the main reason why we have
considered it here.

\section{Discussion}

As we have shown above, the standard formulation of the second
quantized field theory with the background electric fields, which
are capable to create pairs, shows several problems. One of the
main problems among them is the non--cancellation of the IR
divergences. In particular it means that electron self--energy is
IR divergent and there is nothing which can cancel this
divergence. Because the dressing method of Bloch and Nordsiek does
not work in the case if $|in\rangle \neq |out \rangle$. IR
divergent electron self--energy means that electron has an
infinite effective mass.

What kind of conclusions relevant for the back--reaction on the
background fields can we draw out of these observations? Usually
to take into account back--reaction it is tempting to act as
follows \cite{TsamisWoodard}. To find the exact harmonics, define
creation and annihilation operators for them, and then --- define
the vacuum $|in, 0\rangle$ corresponding to the absence of the
positive energy exact in--harmonics. This is supposed to be the
initial state for the problem in question. We should evolve this
state with the use of the exact QED Hamiltonian in the background
field. This way it is tempting to define the rate of the decay of
the background field as follows \cite{TsamisWoodard}. One should
find the evolution of the initial state in question:

\begin{equation} \left|\Psi, t\right\rangle =  {\rm T} e^{-i \int_0^t
\hat{H}_{QED}(E)\, dt} \left|in,0\right\rangle. \end{equation} Or
one could use the functional integral counterpart of this wave
functional. Here $t=0$ is the moment when the constant background
electric field was set up, $t$ is the moment of observation,
$\hat{H}_{QED}(E)$ is the full QED Hamiltonian corresponding to
the exact harmonics, i.e. formulated in the background electric
field $E$. Then one can use this wave functional to find the
created background electric charge whose field compensates the
originally present background electric field \cite{TsamisWoodard}.
This so called Schwinger's in--in formalism in background fields
is applicable only in the case if $E$ is changing slowly in time,
i.e. when there is a slow pair production rate otherwise we can
not use the exact harmonics in fixed background field.

We see that ${\rm T} e^{-i \int_0^t H_{QED}(E)\, dt}$ is a
\emph{non--unitary} evolution operator in the case of the
background field carrying infinite amount of energy. But in
general, even if the background field creates a finite number of
pairs, the QED with the ${\rm T} e^{-i \int_0^t H_{QED}(E)\, dt}$
evolution operator is an ill defined theory due to the problems
with the IR divergences if the time range is taken to infinity.
The Schwinger's method is widely believed to cure out these
problems, because it deals with the finite time range $[0,t]$.

However, our point here is that using the exact harmonics over
background fields in calculations of correlation functions of
\emph{interacting} field theories, one actually deals with
non--closed systems and, as the result, obtains various problems.
In particular the Schwinger's method can not be applied when, the
initial value of $E$ is much bigger than the Schwinger's critical
value and there is a cascade of pair creation. The way out is to
close somehow the system under consideration. How to do that?

For the QED in the background electric field one can do the
following. Let $|0\rangle$ be the Fock vacuum state in QED without
any background fields. To obtain the coherent state which
corresponds to the background field $\vec{E}(x)$, we act on the
vacuum by the shift operator:
\beq
|\vec{E}\rangle=\exp\left[\-i\int{}d^3x
\vec{E}(x)\,\hat{\vec{A}}(x) \right]|0\rangle.
\eeq
Here $\vec{E}(x)$ is the background field whose only non--zero
component is, say, $E_z = E(x)$. It is easy to see that:
\beq
\langle\vec{E}|\hat{F}_{0z}|\vec{E}\rangle=E(x).
\eeq
To find the decay rate of the background field we should find the
evolution of the state $|\vec{E}\rangle$ in time. As the result:
\beq\label{d1}
E(x,t)=\left\langle\vec{E}\left|e^{iH_{QED}\,
t}\hat{F}_{0z}\,e^{-iH_{QED}\, t}\right|\vec{E}\right\rangle,
\eeq
where $H_{QED}$ is the full interacting QED Hamiltonian
\emph{without any} background fields. I.e. one should always
expand around the eventual stable vacuum configuration. The VEV in
question will be calculated elsewhere \cite{Burda}.

We would like to acknowledge discussions with P.Buividovich, S.Gavrilov,
I.Polyubin and P.Burda. The work was partially supported by the
Federal Agency of Atomic Energy of Russian Federation and by the
grant for scientific schools NSh-679.2008.2.

\end{document}